\documentclass[a4paper,12pt]{article}

\usepackage{amsmath,amssymb}
\usepackage{cite}

\begin{document}

\author{Nikolay A. Kudryashov\footnote{E-mail: kudryashov@mephi.ru}}

\title{On Completely Integrability Systems of Differential Equations}

\date{Department of Applied Mathematics, \\ National  Research Nuclear University MEPHI,  \\ 31 Kashirskoe Shosse, 115409, Moscow, \\ Russian Federation}

\maketitle


\begin{abstract}

In this note we discuss the approach which was given by Wazwaz for the proof of the complete integrability to the system of nonlinear differential equations. We show that his method presented in [Wazwaz A.M. Completely integrable coupled KdV and coupled KP systems, Commun Nonlinear Sci Simulat 15 (2010) 2828 -- 2835] is incorrect.

\end{abstract}

\section{Introduction}

In the recent paper \cite{Wazwaz10a} Wazwaz has introduced the approach for the proof of complete integrability to some system of nonlinear differential equations. He believes that the system of nonlinear differential equations is completely integrable if this system has multi soliton solutions.  In \cite{Wazwaz10a} author considered the two following systems of equations: the "new coupled KdV equation"
\begin{equation}\begin{gathered}\label{3}
u_t+u_{xxx}=\frac32(uv)_x+\frac32\,(vw)_x
\\
v_t+v_{xxx}+3\,(w\,u)_x=0,
\\
w_t+w_{xxx}+3\,(u\,v)_x=0
\end{gathered}\end{equation}
and the "new coupled KP system" in the form
\begin{equation}\begin{gathered}\label{4}
(u_t+u_{xxx}-\frac32(uv)_x-\frac32\,(vw)_x)_x+u_{yy}=0,
\\
(v_t+v_{xxx}+3\,(w\,u)_x)_x+v_{yy}=0,
\\
(w_t+w_{xxx}+3\,(u\,v)_x)_x+w_{yy}=0.
\end{gathered}\end{equation}

Author wrote "we aim to apply the Hirota's method to show that the coupled and the coupled systems \eqref{3} and \eqref{4} are completely integrable".

In other work \cite{Wazwaz10b} Wazwaz considered the "coupled KdV equations" in the form
\begin{equation}\begin{gathered}\label{5}
u_t+6\, \alpha u\, u_x- 6\,v\, v_x+\alpha\,u_{xxx}=0,
\\
v_t+3\alpha\,u\,v_x+\alpha\,v_{xxx}=0.
\end{gathered}\end{equation}

Author \cite{Wazwaz10b} says that "the Hirota bilinear method is applied to show this system is completely integrable".

We read in the paper \cite{Wazwaz10c} again "Four M -- component non linear evolution equations, namely the M -- component Korteweg -- de Vries (KdV) equation, the M -- component Kadomtsev -- Petviashvili (KP) equation, the M -- component modified KdV (mKdV) equation and the M -- component mKdV -- KP equation are examined for complete integrability".

From the above mentioned papers one can see that Wazwaz tries to prove the complete integrability of the system of nonlinear differential equations using the Hirota method.

It is well known that there are some definitions for integrability of nonlinear differential equations. The simplified definition of the complete integrability for nonlinear differential equations is the following. Nonlinear evolution equation (NEE) is said to be complete  integrable if NEE can be linearized by some transformation or the Cauchy problem for NEE can be solved by means of the inverse scattering transform.

This definition can be carried across on the system of nonlinear differential equations as well. There are some consequences of the complete integrability for nonlinear evolution equations: the B\"{a}cklund transformations can be found for solutions of NEE; one can obtain the infinite quantities of conservation laws and multi soliton solutions. Let us note that the existence of multi soliton solutions for nonlinear evolution equation is the consequence of the complete integrability.

One of the active participants of our scientific seminar Muraved pointed out in the recent note \cite{Muraved} that  the method by Wazwaz is incorrect for determination of the complete integrability. However Wazwaz did not agree with the arguments by the Muraved note. In the recent comment \cite{Wazwaz10d} he  continues to insist on his point of view.

The aim of this note is to show that the approach by Wazwaz for the proof of complete integrability to system of nonlinear differential equations is not correct. As this takes place we do not discuss the real integrability of the above mentioned systems. We are just going to show that the approach by Wazwaz cannot be used for the proof of the complete integrability. We are going to illustrate that the approach by author \cite{Wazwaz10a, Wazwaz10b, Wazwaz10c}is reduced to the well known solutions of the famous integrable equations. We also demonstrate that solutions for the systems of equations \eqref{3} and \eqref{4} in \cite{Wazwaz10a} are wrong because these solutions do not satisfy the systems of equations \eqref{3} and \eqref{4}.

This note is organized as follows. In section 2 we discuss the method by Wazwaz for the proof of the complete integrability to the systems of nonlinear differential equations. In this section we demonstrate that solutions of systems \eqref{3} and \eqref{4} in \cite{Wazwaz10a}, which were given by author are wrong. In section 3 we formulate the tenth error in finding exact solutions of nonlinear differential equations and present the additional examples which show why we cannot use the approach by Wazwaz for the proof of the complete integrability.

\section{Method by Wazwaz for the proof of the complete integrability}

Let us demonstrate the approach by Wazwaz taking the system of equations \eqref{3} into account.
Author \cite{Wazwaz10a} tried "to show that this system is completely integrable" using "the Hirota's biliniear method".

To prove the complete integrability of the system of equations \eqref{3} he looked for solutions in the form
\begin{equation}\begin{gathered}\label{M1}
u=R_1\,\left(\ln {f}\right)_{xx},\quad
v=R_2\,\left(\ln {f}\right)_{xx},\quad
w=R_3\,\left(\ln {f}\right)_{xx},
\end{gathered}\end{equation}
where $R_1$, $R_2$ and $R_3$ are constants.

Substituting expressions \eqref{M1} into system \eqref{3} author \cite{Wazwaz10a} finds the values of constants $R_1$, $R_2$, $R_3$ and using the linear presentation for equations of system \eqref{3} he obtains the dependence $f$ on $x$ and $t$. After that author \cite{Wazwaz10a} presents some soliton solutions of the system of equations \eqref{3} and makes the conclusion that system \eqref{3} is completele integrable. However this scheme by Wazwaz has the essential defect. Let us show that using constrains \eqref{M1} for solutions of the system \eqref{3} is the basic trouble of Wazwaz's approach.

Let us note that Eq.\eqref{M1} can be written as
\begin{equation}\begin{gathered}\label{M2}
u=R_1\,\left(\ln {f}\right)_{xx},\quad v=\frac{R_2}{R_1}\,u, \quad w=\frac{R_3}{R_1}\,u.
\end{gathered}\end{equation}

Substituting $v$ and $w$ from Eq.\eqref{M2} into Eq.\eqref{3} we reduce the system of equations \eqref{3} to the following
\begin{equation}\begin{gathered}\label{M3a}
u_t+u_{xxx}-\frac{3\,R_2}{2\,R_1}\,(u^2)_x-\frac{3\,R_3\,R_2}{2\,R_1^2}\,(u^2)_x=0,
\end{gathered}\end{equation}
\begin{equation}\begin{gathered}\label{M3b}
u_t+\,u_{xxx}+\frac{3\,R_3}{R_2}\,(u^2)_x=0,
\end{gathered}\end{equation}
\begin{equation}\begin{gathered}\label{M3c}
u_t+u_{xxx}+\frac{3\,R_2}{R_3}\,(u^2)_x=0.
\end{gathered}\end{equation}

Subtracting Eq.\eqref{M3a} from Eq.\eqref{M3b} and Eq.\eqref{M3c} from Eq.\eqref{M3b} we have the system of algebraic equations for $R_1$, $R_2$ and $R_3$ in the form
\begin{equation}\begin{gathered}\label{M4}
\frac{R_3}{R_2}=-\frac{R_1\,R_2+R_3\,R_2}{2\,R_1^2},
\end{gathered}\end{equation}
\begin{equation}\begin{gathered}\label{M5}
\frac{R_3}{R_2}=\frac{R_2}{R_3}.
\end{gathered}\end{equation}

Solving the system of equations \eqref{M4} and \eqref{M5} we obtain
\begin{equation}\begin{gathered}\label{M6}
\frac{R_2}{R_3}=1,\quad \frac{R_2}{R_1}=\frac{-1\pm i \sqrt{7}}{2},
\end{gathered}\end{equation}
\begin{equation}\begin{gathered}\label{M7}
\frac{R_2}{R_3}=-1,\quad \frac{R_2}{R_1}=\frac{1\pm i \sqrt{7}}{2}
\end{gathered}\end{equation}

We obtain that taking into account relations \eqref{M6} and \eqref{M7} (and consequently formulae \eqref{M2} by Wazwaz) the system of equations \eqref{3} is transformed to two following systems of equations
\begin{equation}\begin{gathered}\label{M8}
u_t+{6}{}\,u\,u_x+u_{xxx}=0, \quad v=\frac{R_2}{R_1}\,u, \quad w=\frac{R_3}{R_1}\,u.
\end{gathered}\end{equation}
and
\begin{equation}\begin{gathered}\label{M9}
u_t-{6}{}\,u\,u_x+u_{xxx}=0, \quad v=\frac{R_2}{R_1}\,u, \quad w=\frac{R_3}{R_1}\,u.
\end{gathered}\end{equation}

It is clear that the systems of equations \eqref{M8} and \eqref{M9} are not equivalent to the system \eqref{3}. Note that the first equation of the systems of equations \eqref{M8} and \eqref{M9} is the famous Korteweg - de Vries equation \cite{Korteweg, Kruskal, Ablowitz, Kudryashov}. There are soliton solutions of this equation \cite{Gardner, Lax} and these solutions can be found by the well known Hirota method \cite{Hirota}. The second and the third equations are trivial algebraic formulae for finding $v$ and $w$. No doubt that the systems of equations \eqref{M8} and \eqref{M9} are complete integrable system but this system is not equivalent to the system \eqref{3}.

So, we obtain that using constrains for solutions of the system of equations \eqref{3} author \cite{Wazwaz10a} transformed the origin system of equations to another system of equations and proved the complete integrability for the new system of equations \eqref{M8} and \eqref{M9}. Author did not study system \eqref{3} but considered systems \eqref{M8} and \eqref{M9}.

Soliton solutions of the system of equations \eqref{M9} are found by formulae

\begin{equation}\begin{gathered}\label{M9a}
u=2\,\frac{\partial^2 \ln{F}}{\partial x^2}, \quad v=\left(-1\pm{i\sqrt{7}}{}\right)\frac{\partial^2 \ln{F}}{\partial x^2}, \quad w=\left(-1\pm{i\sqrt{7}}{}\right) \frac{\partial^2 \ln{F}}{\partial x^2},
\end{gathered}\end{equation}
and
\begin{equation}\begin{gathered}\label{M9b}
u=-2\,\frac{\partial^2 \ln{F}}{\partial x^2}, \quad v=-\left(1\pm{i\sqrt{7}}{}\right)\frac{\partial^2 \ln{F}}{\partial x^2}, \quad w=\left(1\pm{i\sqrt{7}}{}\right) \frac{\partial^2 \ln{F}}{\partial x^2}.
\end{gathered}\end{equation}
We have one soliton solution if we take in formulae \eqref{M9a} and \eqref{M9b}
\[F=F_1=1+e^{k\,x-k^3\,t-k\,x_0}.\]

Taking  into consideration $F(x,t)=F_2$, where
\[F_2=1+e^{\theta_1}+e^{\theta_2}+e^{\theta_1+\theta_2+A_{12}}, \quad \theta_i=k_i\,x-k_i^3\,t-k\,x_0^{(i)},\quad e^{A_{12}}=\frac{(k_1-k_2)^2}{(k_1+k_2)^2}\]
we obtain two soliton solutions of system \eqref{3} using formulae \eqref{M9a} and \eqref{M9b} and so on.

Author \cite{Wazwaz10a} has given the following values for the parameters $R_1$, $R_2$
and $R_3$ in the work \cite{Wazwaz10a}:
\begin{equation}\begin{gathered}\label{WWW}
R_1=\pm 2,\quad R_2=4,\quad R_3=\mp\,4.
\end{gathered}\end{equation}
However his solutions do not satisfy the system of equations \eqref{3} and consequently his solutions are wrong.

Author \cite{Wazwaz10a} applied his approach to the system \eqref{4}  as well. He again looked for solutions in the form
\begin{equation}\begin{gathered}\label{M1b}
u=R_1\,\left(\ln {f}\right)_{xx},\quad
v=R_2\,\left(\ln {f}\right)_{xx},\quad
w=R_3\,\left(\ln {f}\right)_{xx},
\end{gathered}\end{equation}
where $R_1$, $R_2$ and $R_3$ are constants as well. We have from \eqref{M1b} equalities
\begin{equation}\begin{gathered}\label{M2b}
v=\frac{R_2}{R_1}\,u\quad
w=\frac{R_3}{R_1}\,u.
\end{gathered}\end{equation}

Substituting \eqref{M2b} into system of equations \eqref{4} we obtain
\begin{equation}\begin{gathered}\label{M4ba}
\left(u_t+u_{xxx}-\left(\frac{3\,R_2}{\,R_1}+\frac{3\,R_3\,R_2}{\,R_1^2}\right)\,u\,u_x\right)_x+u_{yy}=0,
\end{gathered}\end{equation}
\begin{equation}\begin{gathered}\label{M4bb}
(u_t+u_{xxx}+6\,\frac{R_3}{R_2}\,u\,u_x)_x+u_{yy}=0,
\end{gathered}\end{equation}
\begin{equation}\begin{gathered}\label{M4bc}
(u_t+u_{xxx}+6\frac{\,R_2}{R_3}\,u\,u_x)_x+w_{yy}=0.
\end{gathered}\end{equation}

Subtracting \eqref{M4ba} from \eqref{M4bb} and \eqref{M4bc} from \eqref{M4bb} we have the system of algebraic equations for constants $R_1$, $R_2$ and $R_3$ in the form
\begin{equation}\begin{gathered}\label{M5b}
-\frac{R_2}{\,R_1}-\frac{R_3\,R_2}{\,R_1^2}=2\,\frac{R_3}{R_2},\quad \frac{\,R_2}{R_3}=\,\frac{R_3}{R_2}.
\end{gathered}\end{equation}

Solving the system \eqref{M5b} we have
\begin{equation}\begin{gathered}\label{M6ba}
\frac{R_3}{R_2}=1,\quad \frac{R_2}{R_1}=-\frac12\pm \frac{i\,\sqrt{7}}{2}
\end{gathered}\end{equation}
and
\begin{equation}\begin{gathered}\label{M6bb}
\frac{R_3}{R_2}=-1,\quad \frac{R_2}{R_1}=\frac12\pm \frac{i\,\sqrt{7}}{2}.
\end{gathered}\end{equation}

We obtain the result as in previous example. Assuming constrains \eqref{M1b} author \cite{Wazwaz10a}  solved the following system of equations
\begin{equation}\begin{gathered}\label{M7b}
(u_t+u_{xxx}\pm 6\,\,u\,u_x)_x+u_{yy}=0,\quad v=\frac{R_2}{R_1}\,u, \quad
w=\frac{R_3}{R_1}\,u,
\end{gathered}\end{equation}
where expressions $\frac{R_2}{R_1}$ and $\frac{R_3}{R_1}$ are determined by formulae \eqref{M6ba} and \eqref{M6bb}.
The first equation in \eqref{M7b} is the famous Kadomtsev -- Petviashvily equation. Soliton solutions of this equation are well known.
Solutions of Eq. \eqref{M7b} take the form

\begin{equation}\begin{gathered}\label{M9a}
u=2\,\frac{\partial^2 \ln{F}}{\partial x^2}, \quad v=\left(-1\pm{i\sqrt{7}}{}\right)\frac{\partial^2 \ln{F}}{\partial x^2}, \quad w=\left(-1\pm{i\sqrt{7}}{}\right) \frac{\partial^2 \ln{F}}{\partial x^2}
\end{gathered}\end{equation}
and
\begin{equation}\begin{gathered}\label{M9b}
u=-2\,\frac{\partial^2 \ln{F}}{\partial x^2}, \quad v=-\left(1\pm{i\sqrt{7}}{}\right)\frac{\partial^2 \ln{F}}{\partial x^2}, \quad w=\left(1\pm{i\sqrt{7}}{}\right) \frac{\partial^2 \ln{F}}{\partial x^2},
\end{gathered}\end{equation}
We have one soliton solution if we take in formulae \eqref{M9a} and \eqref{M9b}
\[F=F_1=1+e^{k_1\,x+k1\,m_1\,y-k_1^3\,t -k1\,m1^2\,t-k_1\,x_0}.\]

Multi soliton solutions are found using the well known formulae \cite{Satsuma}. Taking  into consideration $F(x,y,t)=F_2$, where
\begin{equation}\begin{gathered}\label{M99b}
F_2=1+e^{\theta_1}+e^{\theta_2}+e^{\theta_1+\theta_2+A_{12}}, \quad \theta_i=k_i\,x+k_i\,r_i\,y-k_i^3\,t-k_i\,r_i^2\,t-k_i\,x_0^{(i)},\\
\\
(i=1,2), \quad e^{A_{12}}=\frac{3(k_1-k_2)^2-(r_1-r_2)^2}{3\,(k_1+k_2)^2-(r_1-\,r_2)^2}
\end{gathered}\end{equation}
we obtain two soliton solutions of system \eqref{3} from formulae \eqref{M9a} and \eqref{M9b} and so on.

Author \cite{Wazwaz10a}  gave the following values for the parameters $R_1$, $R_2$
and $R_3$ :
\begin{equation}\begin{gathered}\label{WWW}
R_1=\pm 2,\quad R_2=4,\quad R_3=\mp\,4,
\end{gathered}\end{equation}
but  his solutions do not satisfy the system of equations \eqref{3}. His solutions are wrong again.

Let us shortly consider the application by Wazwaz for the proof of the complete integrability to the system of equations \eqref{5}. This example was considered in the note \cite{Muraved} but Wazwaz in his comment \cite{Wazwaz10d} did not agree with arguments by Muraved.

The author \cite{Wazwaz10d} looked for the solutions of the equation system \eqref{5} in the form
\begin{equation}\begin{gathered}\label{Mur2}
u=R\,\left(\ln {f}\right)_{xx},\quad
v=R_1\,\left(\ln {f}\right)_{xx},
\end{gathered}\end{equation}
where $R$ are $R_1$ are constants.

It is obviously that taking constrains \eqref{Mur2} for solutions to the system of equations \eqref{5} we obtain the following relation between variables $u$ and $v$
\begin{equation}\begin{gathered}\label{Mur3}
v=\frac{R_1}{R}u.
\end{gathered}\end{equation}

Taking \eqref{Mur3} into account we can reduce the system of Eq.\eqref{5} to the following
system
\begin{equation}\begin{gathered}\label{Mur4a}
u_t+6\,\alpha\,u\,u_x-6\frac{R_1^2}{R^2}\,u\,u_x+\alpha\,u_{xxx}=0
\end{gathered}\end{equation}
\begin{equation}\begin{gathered}\label{Mur4b}
u_t+3\,\alpha\,u\,u_x+\alpha\,u_{xxx}=0,
\end{gathered}\end{equation}

Subtracting Eq.\eqref{Mur4b} from Eq.\eqref{Mur4a} we have
\begin{equation}\begin{gathered}\label{Mur5}
\left(3\,\alpha-6\,\frac{R_1^2}{R^2}\right)\,u\,u_x=0.
\end{gathered}\end{equation}

Taking into account that $u\,u_x\neq0$ we obtain from \eqref{Mur5} the relation between $R_1$ and $R$ in the form
\begin{equation}\begin{gathered}\label{Mur6}
\frac{R_1}{R}=\pm\frac{\sqrt{2\alpha}}{2}.
\end{gathered}\end{equation}

So we obtain the result similar to the first example. Using constrains for solutions to the system of equations  \eqref{5} we have the system of equations in the form
\begin{equation}\begin{gathered}\label{Mur7}
u_t+3\,\alpha\,u\,u_x+\alpha\,u_{xxx}=0, \quad v=\frac{R_1}{R}\,u,
\end{gathered}\end{equation}
where $\frac{R_1}{R}$ is determined by formula \eqref{Mur6}. If we use transformations $u={2}\,{u^{'}}$ and $t=\frac {t^{'}}{\alpha}$ we will obtain the standard form of the KdV equation. Multi soliton solutions of the KdV equation were found in 1971 by Hirota \cite{Hirota}. Soliton solutions of the first equation in \eqref{Mur7} can be found by formula
\begin{equation}\begin{gathered}\label{Mur8}
u=4\,\frac{\partial^2 \ln{F}}{\partial x^2}.
\end{gathered}\end{equation}

These results by Muraved are absolutely right and I agree with him completely. Small misprint in
formula (9) \cite{Muraved} did not follow from previous expressions by Muraved.

\section{Additional examples and the tenth error in finding exact solutions of nonlinear differential equations}

In this section we present some additional examples and formulate the tenth error in finding exact solutions for the systems of nonlinear differential equations.

Let us start using the trivial linear system of ordinary differential equations to demonstrate the defect of approach which was given by Wazwaz.

\emph{Example 1}.
Consider the system of differential equations
\begin{equation}\begin{gathered}\label{S1a}
u_z+u=v
\end{gathered}\end{equation}
\begin{equation}\begin{gathered}\label{S1b}
v_z+v=u
\end{gathered}\end{equation}

Using the constraint by Wazwaz we take $v=u$. As the result we have the differential equation in the form
\begin{equation}\begin{gathered}\label{S2}
u_z=0.
\end{gathered}\end{equation}

The general solution of this equation is the constant
\begin{equation}\begin{gathered}\label{S2}
u=v=C_1.
\end{gathered}\end{equation}

However substituting $v=u_z+u$ into Eq.\eqref{S1b} we
obtain the equation in the form
\begin{equation}\begin{gathered}\label{S3}
u_{zz}+2\,u_{z}=0.
\end{gathered}\end{equation}

The general solution of Eq.\eqref{S3} can be presented in the form
\begin{equation}\begin{gathered}\label{S4}
u(z)=C_1-\frac{C_2}{2}\,e^{-2z},
\end{gathered}\end{equation}
where $C_1$ and $C_2$ are arbitrary constants.

The system of equations \eqref{S1a} and \eqref{S1b} is integrable because there is the general solution of this system in the form
\begin{equation}\begin{gathered}\label{S4}
u(z)=C_1-\frac{C_2}{2}\,e^{-2z}, \quad v=C_1+\frac{\,C_2}{2}e^{-2\,z}.
\end{gathered}\end{equation}

We see that solutions \eqref{S2} and \eqref{S4} of the system of equations \eqref{S1a} and \eqref{S1b} are different.

\emph{Example 2}. Consider the system of nonlinear differential equations
\begin{equation}\begin{gathered}\label{S5a}
u_z+u^2=v
\end{gathered}\end{equation}
\begin{equation}\begin{gathered}\label{S5b}
v_z+v^2=u
\end{gathered}\end{equation}

Using the constraint by Wazwaz we take $v=u$ for solutions of the system of equations \eqref{S5a} and \eqref{S5b}. As the result we have only one differential equation in the form
\begin{equation}\begin{gathered}\label{S6}
u_z+u^2-u=0.
\end{gathered}\end{equation}

Exact solution of this equation takes the form
\begin{equation}\begin{gathered}\label{S7}
u=\frac{1}{1+C_1\,e^{-z}}.
\end{gathered}\end{equation}

However substituting $v=u_z+u^2$ from Eq.\eqref{S5a} into Eq.\eqref{S5b} we
obtain equation
\begin{equation}\begin{gathered}\label{S8}
u_{zz}+u_z^2+2\,u\,u_{z}+2\,u^2\,u_z+u^4-u=0.
\end{gathered}\end{equation}

We do not know the general solution of Eq.\eqref{S8}. Special solution of this equation does not allow to hope that the system of equations \eqref{S5a} and \eqref{S5b} is integrable.
So the method by Wazwaz does not allow to determine the complete integrability for the system of nonlinear ordinary differential equations.

\emph{Example 3}. Consider the Henon -- Heiles system
\begin{equation}\begin{gathered}\label{S9a}
\ddot{x}=-\alpha\,x-2\,\delta\,x\,y,\quad \ddot{x}\equiv \frac{d^x}{dt^2},
\end{gathered}\end{equation}
\begin{equation}\begin{gathered}\label{S9b}
\ddot{y}=-\beta \,y +\gamma\,y^2 - \delta\,x^2, \quad \ddot{y}\equiv \frac{d^2y}{dt^2}.
\end{gathered}\end{equation}

Using the approach by Wazwaz we look for solutions of the system of equations \eqref{S9a} and \eqref{S9b} in the form
\begin{equation}\begin{gathered}\label{S9bb}
y=A\,x.
\end{gathered}\end{equation}

Substituting the constrain of solution  \eqref{S9b} into the system of equations \eqref{S9a} and \eqref{S9b} we have that Eqs.\eqref{S9a}, \eqref{S9b} are transformed at $\alpha=\beta$, $A=\pm \sqrt{\frac{\delta}{\gamma+2\,\delta}}$ to the equation in the form
\begin{equation}\begin{gathered}\label{S10}
\ddot{x}=-\alpha\,x \mp \,\sqrt{\frac{4\,\delta^3}{\gamma+2\,\delta}}\,\,x^2.
\end{gathered}\end{equation}

Multiplying \eqref{S10} on $\dot{x}$ and integrating with respect to $t$ we have equation
\begin{equation}\begin{gathered}\label{S11}
\dot{x}^2=-\alpha\,x^2 \mp \,\frac43\,\sqrt{\frac{\delta^3}{\gamma+2\,\delta}}\,\,x^3+C_1,
\end{gathered}\end{equation}
where $C_1$ is a constant of integration.
The general solution of \eqref{S11} is expressed via the Weierstrass elliptic function but we know exactly that the Henon -- Heilis system is not integrable in the general case.

This example has showed again that the special solutions cannot be proof for the complete integrability of nonlinear differential equations and the approach by Wazwaz for proof of the complete integrability is not correct.

We considered systems of nonlinear differential equations but now we are going to present the example with a partial differential equation which is not integrable.

\emph{Example 4}. Consider the equation in the form
\begin{equation}\begin{gathered}\label{K1a}
u_{xt}+u\,u_t+6\,u^2\,u_x+6\,u_{x}^2+6\,u\,u_{xx}+u\,u_{xxx}+u_{xxxx}=0.
\end{gathered}\end{equation}

Eq.\eqref{K1a} is not completely integrable because we do not have the Lax pair for this equation. However this equation can be written as

\begin{equation}\begin{gathered}\label{K1aa}
\left(\frac{\partial }{\partial x} +u \right) \left(u_t+6\,u\,u_x+u_{xxx} \right)=0
\end{gathered}\end{equation}

From Eq.\eqref{K1aa} one can see that there are multi solitary wave solutions of this equation because all multi soliton solutions of the KdV equation are solutions of Eq.\eqref{K1a} but Eq. \eqref{K1a} is not complete integrable and we cannot solve the Cauchy problem for this equation.

In review \cite{Kudryashov09c} we have presented the list of the seven errors that we observed in many works devoted for finding exact solutions of nonlinear differential equations. These errors were also discussed in other works \cite{Kudryashov09a, Kudryashov09b, Kudryashov09d, Kudryashov09e, Kudryashov10a, Kudryashov10b, Kudryashov10c, Kudryashov10d, Kudryashov10e, Disine, Parkes09a, Parkes09b, Parkes09c, Parkes10a, Parkes10b}. Authors \cite{Popovich} extended our list adding two other errors. Now we add the tenth error to this list.

\emph{\textbf{Tenth error}}. \emph{Studying the system of nonlinear differential equations some authors do not take into consideration that application of the anzatz methods to the system of equations leads to the nonequivalent modification of the origin system of equations.}

In essence these authors look for exact solutions of other systems of equations. Let us explain our point of view. Assume that we have the system of nonlinear ordinary differential equations in the form
\begin{equation}\begin{gathered}\label{W1}
E_1(u,v,u_z,v_z, \ldots)=0\\
E_2(u,v,u_z,v_z, \ldots)=0.
\end{gathered}\end{equation}
Let us also suggest that we can find some exact solutions of this system using the tanh -- function method
\begin{equation}\begin{gathered}\label{W2}
u=A_1+B_1\,\tanh{kz},\\
v=A_2+B_2\,\tanh{k\,z},
\end{gathered}\end{equation}
where $A_1$, $B_1$, $A_2$ and $B_2$ are constants that are determined after substitution \eqref{W2} into the system of equations \eqref{W1}.

However it is easy to see from expressions \eqref{W2} that
\begin{equation}\begin{gathered}\label{W3}
u=A_1-\frac{B_1\,A_2}{B_2}+\frac{B_1}{B_2}\,v.
\end{gathered}\end{equation}

We have to remember that substituting \eqref{W3} into the system of equations \eqref{W1} we change the origin system of equations. This system is reduced to the ordinary differential equation and to the trivial relation \eqref{W3}. This system is not equivalent to the origin system of equations. In essence  using \eqref{W3} we look for exact solutions of the ordinary differential equations. Unfortunately some authors do not recognize this fact and do not take this error into consideration.

\end{document}